\documentclass[journal]{IEEEtran}

\usepackage{amsmath}
\usepackage{graphicx}
\usepackage{caption}
\usepackage{url}
 \usepackage{multirow}
\graphicspath{{Figures/}}
\usepackage{subcaption}
\usepackage[sort,compress]{cite}
\usepackage{times}
\usepackage{algorithm}
\usepackage{verbatim}
\usepackage{sidecap}
\usepackage{array}
\usepackage{float}
\usepackage{comment}
\usepackage{minibox}
\usepackage{color,soul}
\usepackage{enumerate}

\usepackage{xcolor,soul}
\colorlet{shadecolor}{yellow}
\usepackage{mathtools}
\usepackage{multicol}
\usepackage{dblfloatfix}
\input epsf

\begin{document}

\title{XDWM: A 2D Domain Wall Memory}

\author{Arifa Hoque, Alex K. Jones, and Sanjukta Bhanja}

\maketitle

\begin{abstract}

\bf{Domain-Wall Memory (DWM) structures typically bundle nanowires shifted together for parallel access.  Ironically, this organization does not allow the natural shifting of DWM to realize \textit{logical shifting} within data elements.  We describe a novel 2-D DWM cross-point (X-Cell) that allows two individual nanowires placed orthogonally to share the X-Cell.  Each nanowire can operate independently while sharing the value at the X-Cell.  Using X-Cells, we propose an orthogonal nanowire in the Y dimension overlaid on a bundle of X dimension nanowires for a cross-DWM or XDWM.  We demonstrate that the bundle shifts correctly in the X-Direction, and that data can be logically shifted in the Y-direction providing novel data movement and supporting processing-in-memory.  
We conducted studies on the requirements for physical cell dimensions and shift currents for XDWM. Due to the non-standard domain, our micro-magnetic studies demonstrate that XDWM introduces a shift current penalty of 6.25\% while shifting happens in one nanowire compared to a standard nanowire.  We also demonstrate correct shifting using nanowire bundles in both the X- and Y- dimensions.  Using magnetic simulation to derive the values for SPICE simulation we show the maximum leakage current between nanowires when shifting the bundle together is $\le3\%$ indicating that sneak paths are not problematic for XDWM.} 
\end{abstract}
\vspace*{-0.2in}
\section{Introduction}
\label{sec:intro}

\begin{figure}[h]
\center 
`\includegraphics[width=3.4in]{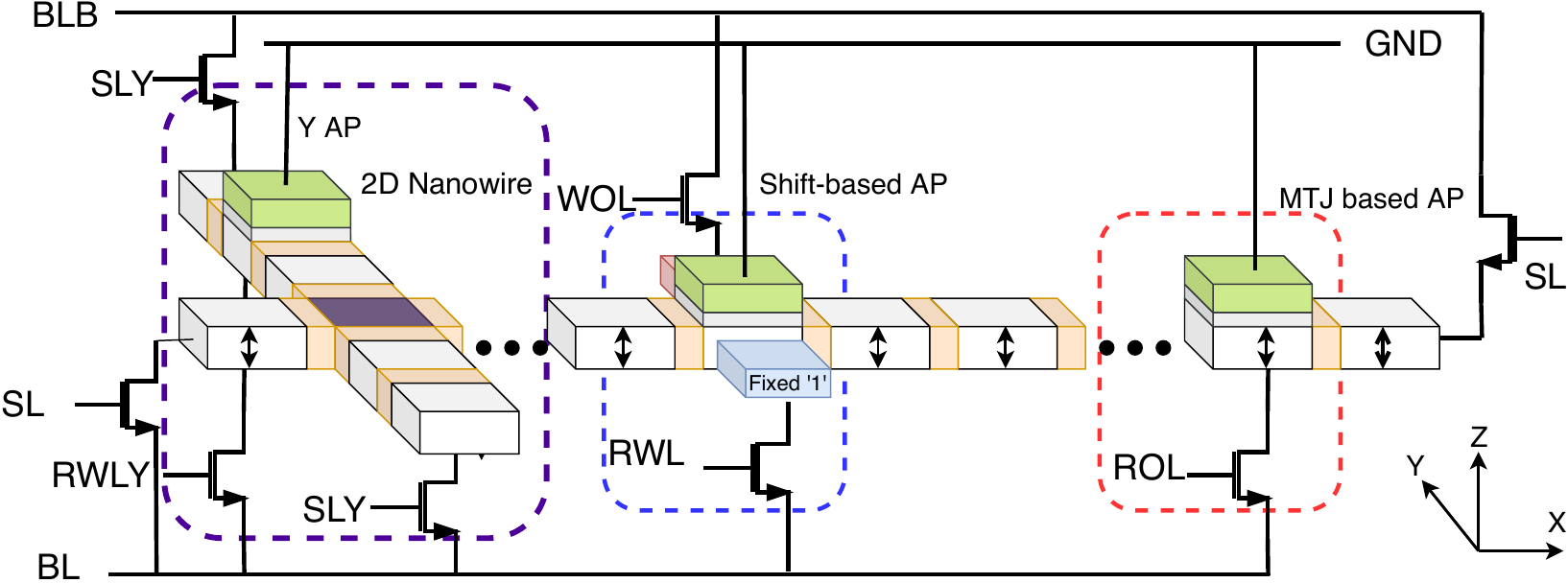}
\caption{2-D domain wall memory concept with X/Y-Dimension nanowires overlaid.  Each nanowire has the access ports for read, write and shift operations.}
\label{2-D DWM cell}
\vspace*{-0.2 in}
\end{figure}

Domain-Wall Memory (DWM), also known as ``Racetrack'' memory~\cite{parkin2008magnetic,deger2020multibit,khan2019shiftsreduce} is akin to spin transfer torque- magnetic random access memory (STT-MRAM) but expands the free layer of the magnetic tunnel junction (MTJ) into a ferromagnetic nanowire.  As such, DWM inherits many positive STT-MRAM attributes including SRAM-class access speed, reduced static power, high endurance, CMOS compatibility, etc.  DWM nanowires consist of many magnetic domains separated by domain walls at patterned notches that form ``pinning sites'' in the nanowire, represented in Fig.~\ref{2-D DWM cell}.  Each domain has its own magnetization direction based on either in-plane (+X/-X) or perpendicular (+Z/-Z) magnetic anisotropy (IMA or PMA).  Binary values are represented by the magnetization direction of each domain, either parallel or antiparallel to a fixed reference (green). If two adjacent domains have opposite magnetization a domain-wall (DW), \textit{i.e.,} a gradual reorientation of magnetic moments, is formed at the pinning site.  Several domains share an access point (AP) for read and write operations. 

DW motion occurs when a sufficient current is applied (\textit{i.e.,} from \texttt{SL}, \texttt{GND}$=$`Z') to counterbalance the pinning potential and depin the walls~\cite{hayashi2008current} to shift along the nanowire. To read or write to a domain, the domains are shifted to the MTJ-based APs. 
\texttt{ROL} allows current from \texttt{BL}$\rightarrow$\texttt{GND} for read/write (red dashed box).  Shift-based APs  (blue dashed box) activate \texttt{RWL} and \texttt{WOL} (\texttt{GND}$=$`Z') to shift orthogonally from magnetically fixed fins with the direction (storing 1/0) controlled by \texttt{BL}/$\overline{\text{\texttt{BL}}}$.  This improves write efficiency over current-base writes~\cite{venkatesan2013dwm}.

We propose cross-DWM (XDWM) to add a Y-direction nanowire (purple dashed box) with its own shifting control (\texttt{SLY}) and potentially dedicated APs (\texttt{RWLY}).  
Each orthogonal nanowire can operate individually but with one shared domain at the nanowire cross-point (X-Cell) (purple). Typically, one (or more) Y nanowires (Y-NWs) will be overlaid with a ``bundle'' of X nanowires (X-NWs) that shift together and comprise memory rows~\cite{khan2019shiftsreduce}.  XDWM allows data to be shifted orthogonally from one nanowire (\textit{i.e.,} bit position) to another, enabling logical shifts without expensive peripheral circuitry, which complements other processing in DWM proposals~\cite{yu2014energy,ollivier2021pirm}. Thus, to demonstrate XDWM feasibility, we analyze X-Cell magnetization to check the stability of the domain wall formed in both dimensions of the crossed nanowires.  

\vspace{-.1in}
\section{Proposed 2-D Domain Wall Memory Design}
\label{sec:design}
\begin{figure}
\begin{center}
\begin{tabular}{c c}
\begin{tabular}{c}
\includegraphics[scale=0.25]{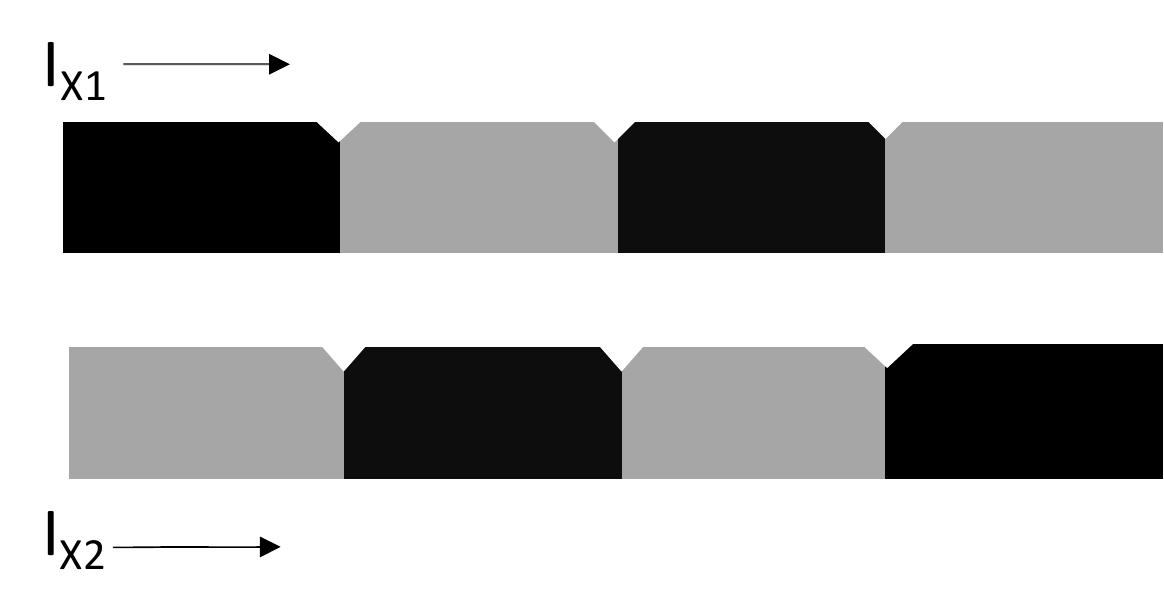}\\(a) \\
\includegraphics[scale=0.25]{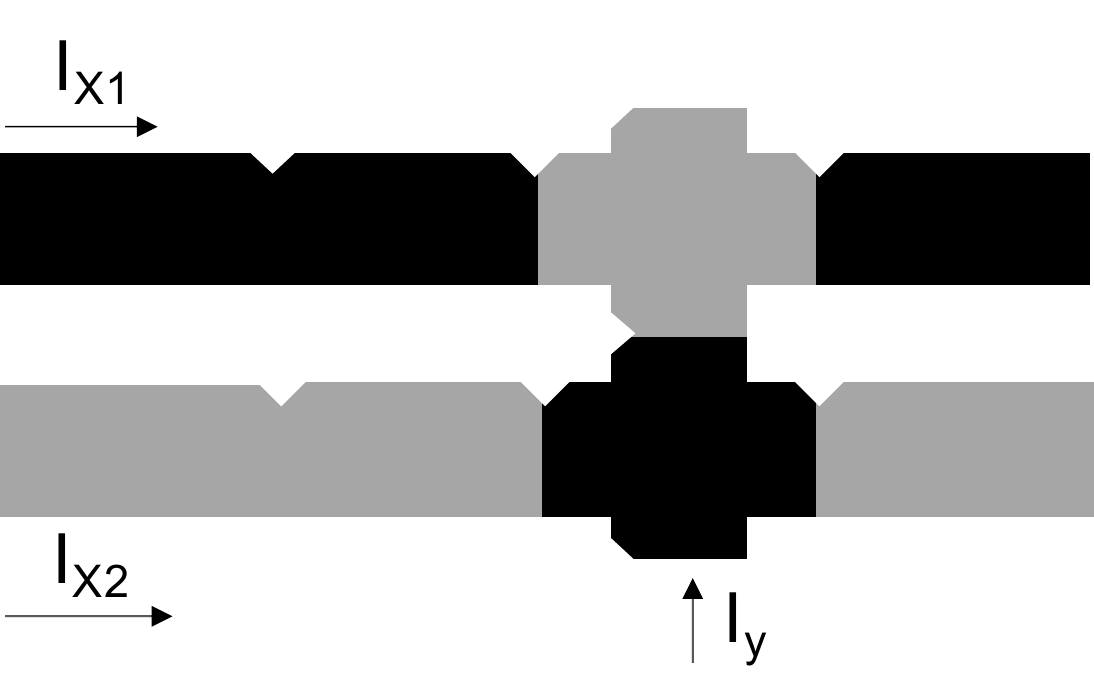}\\(b) 
\end{tabular}
&
\hspace*{-0.2in}
\begin{tabular}{c}
\includegraphics[scale=0.27]{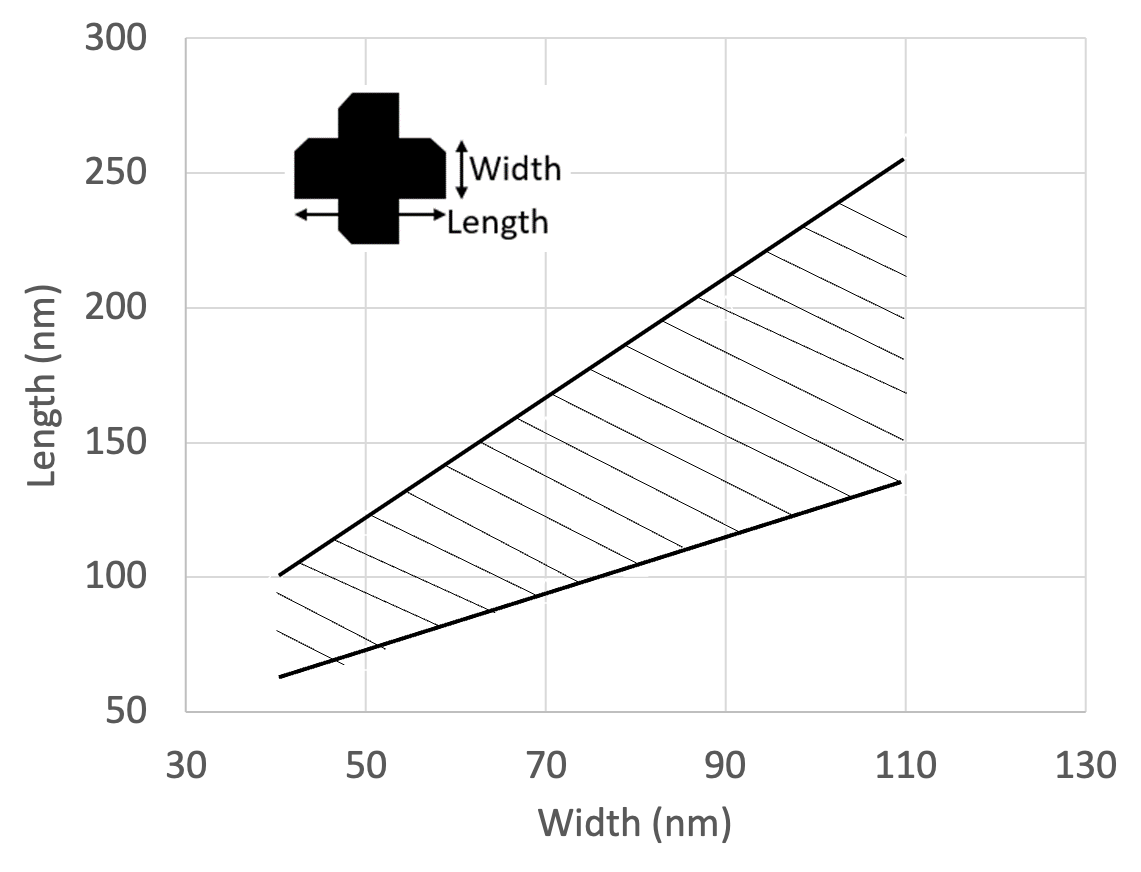}\\
(c)
\end{tabular}
\end{tabular}
\end{center}
\vspace*{-0.15in}\caption{(a) Conventional X-Nanowires (X-NW) and  (b) 2D (XDWM) nanowire and (c) Dimension analysis of the X-cell}
\label{fig:XDWM}
\end{figure}

Fig.~\ref{fig:XDWM}(a) shows a comparative illustration of a conventional nanowire bundle and the XDWM structure for a few domains in Fig.~\ref{fig:XDWM}(b).  
Note, while we assume planar devices to avoid the complications of additional process steps to realize 3D DWM~\cite{parkin2015memory} conceptually XDWM could be applied to 3D devices. At the cross-points, we can conceptualize the shared domain as domain in one dimension with ``fins'' in the other; \textit{i.e.,} from the X perspective, the shared domain has fins in the Y dimension.  The magnetization of the cross nanowire does not affect the DW motion in either direction. However, the DW velocity and shifting currents in X-NW and Y-NW depend on the number of cross domains, which is discussed in Section~\ref{sec:results}. Next, we provide some preliminary details about our material and magnetic anisotropy assumptions of the XDWM nanowires.

 {\bf Magnetic Anisotropy:} To allow shifting in both the X and Y dimension in the same device, it is necessary to use PMA.  The parallel torque to +Z/-Z can be applied in either X or Y dimension.  Attempting this with IMA, +X/-X magnetization can only function in the X-direction.  Y polarized shifting current cannot apply torque on +X/-X oriented domains.  It would be necessary to turn the orientation from +X/-X to +Y/-Y at the X-Cell, which is a much more dubious and expensive prospect. Moreover, most STT and DWM technologies rely on material with PMA for better density and energy-efficient write and shift operations. Furthermore, PMA achieves better DW velocity compared to IMA-nanowires. 

{\bf X-Cell shape:} As domain length is longer than width for efficient DW motion, we considered making the Y-NW wider than its length.  The goal was to avoid penalizing the performance and energy of shifting in the X-NW bundle while accepting an inefficient shifting operation in the Y-NW.  Unfortunately, this device was insufficiently stable.  Ultimately, we obtained our best results from a ``cross-shaped'' structure of two overlaid rectangles.  The fins from this unique shape affect the shape anisotropy and demagnetization field.  Thus, for successful operation of the X-cell, a stable single domain magnetization at the cross-domain is necessary.

{\bf Dimension Study:} The width of the X-cell is constrained due to the requirement of stable single domain magnetization in the cross-domain. We performed a geometric variation study to determine the range of dimensions at which the XDWM is functional. We varied the width of the X-cell, \textit{i.e.,} the length of the Y-NW, from 40 to 110nm while sweeping the length of the X-cell from 50 to 300nm and observe the magnetization of the cross-domain. The area enclosed by the two lines of Fig.~\ref{fig:XDWM}(c) expresses the acceptable range of dimensions of the cross-domain to form and maintain single domain magnetization. 
Dimensions of both X- and Y-NWs are selected to be 80 and 40nm, with expectations of low shift current and stable magnetization. To combat process variation however, the dimensions might be in the middle region of the area shown in Fig.~\ref{fig:XDWM}(b). 

\vspace*{-0.1in}
\section{Simulation Framework}

The shift current density ($J_{shift}$) is calculated by the Landau-Lifshitz-Gilbert (LLG) equation having adiabatic and non-adiabatic torques~\cite{hayashi2008current} (assuming the current flows in the $\mathbf{x}$-direction) as described in Eq.~\ref{LLG}:
\vspace{-.05in}
\begin{equation}
\frac{d\mathbf{m}}{dt}=-\gamma\mathbf{m}\times\mathbf{H}_{eff}+\alpha\mathbf{m}\times\frac{d\mathbf{m}}{dt}-|\mathbf{u}|\frac{\partial\mathbf{m}}{\partial x}+ \beta|\mathbf{u}|\mathbf{m}\times\frac{\partial\mathbf{m}}{\partial x}
\label{LLG}
\end{equation}
\noindent
$$
\text{such that  } \hspace{0.02in} \mathbf{H}_{eff}= \mathbf{H}_{ex}+\mathbf{H}_{d}+\mathbf{H}_{anis}, \hspace{0.02in} \mathbf{u}=\frac{\mu_B\mathbf{J}P}{2eM_s}
$$

\noindent where $\gamma$, $\alpha$ and $\beta$ are the gyromagnetic ratio, Gilbert damping constant, and non-adiabatic torque coefficient. $\mathbf{u}$ is velocity in the direction of current, and depends on the Bohr magneton ($\mu_B$), polarization (P), electronic charge (e), saturation magnetization ($M_s$) along with the flowing current density ($J$). The effective field ($\mathbf{H}_{eff}$) is a vector summation of exchange ($\mathbf{H}_{ex}$), demagnetization ($\mathbf{H}_{d}$) and anistropic fields ($\mathbf{H}_{anis}$).


Additional simulation parameters are: Exchange stiffness $A=1.0\cdot10^{-11}J/m$; damping constant $\alpha =0.02$  based on conventional PMA materials such as Co, CoFe or CoFeB alloys. Saturation Magnetization $M_{s} = 6\cdot10^5A/m$ and $K_{u} = 0.59\cdot10^6 J/m^{3}$, STT non-adiabatic parameter $\beta = 0.04$, and spin polarization $P = 0.72$. We have utilized a GPU-based  micromagnetic simulator, \textit{Mumax3}~\cite{vansteenkiste2014design} with a cell size of 2nm$\times$2nm$\times$1nm. The polarized electric current was fed along the nanowire in the +X-direction. 

{\bf Resistance Modeling:} We used a finite difference solver (LLG) to calculate resistance values for different combinations of stored data. For a X-directed current  flow, the nanowire is divided into a grid of cells having dimension of $dx$,$dy$, and $dz$ each. Our model calculates the anisotropic magneto-resistance (AMR) as $AMRc\times(\mathbf{m_{1}}*\mathbf{m_{2}})$ where $\mathbf{m_{1}}$ and $\mathbf{m_{2}}$ are  two neighbor cells. We used $AMR_c$ of 0.014~\cite{ruffer2014anisotropic} in the simulation. The resistance of each cell is then calculated by $r=\rho \frac{dx}{dydz}\times(1+AMR)$.   The total resistance of the nanowire is the series-parallel combination of all cells' resistances. 

\section{Results and Discussion}
\label{sec:results}

\begin{figure}[h]
\center     
\includegraphics[width=3.0in]{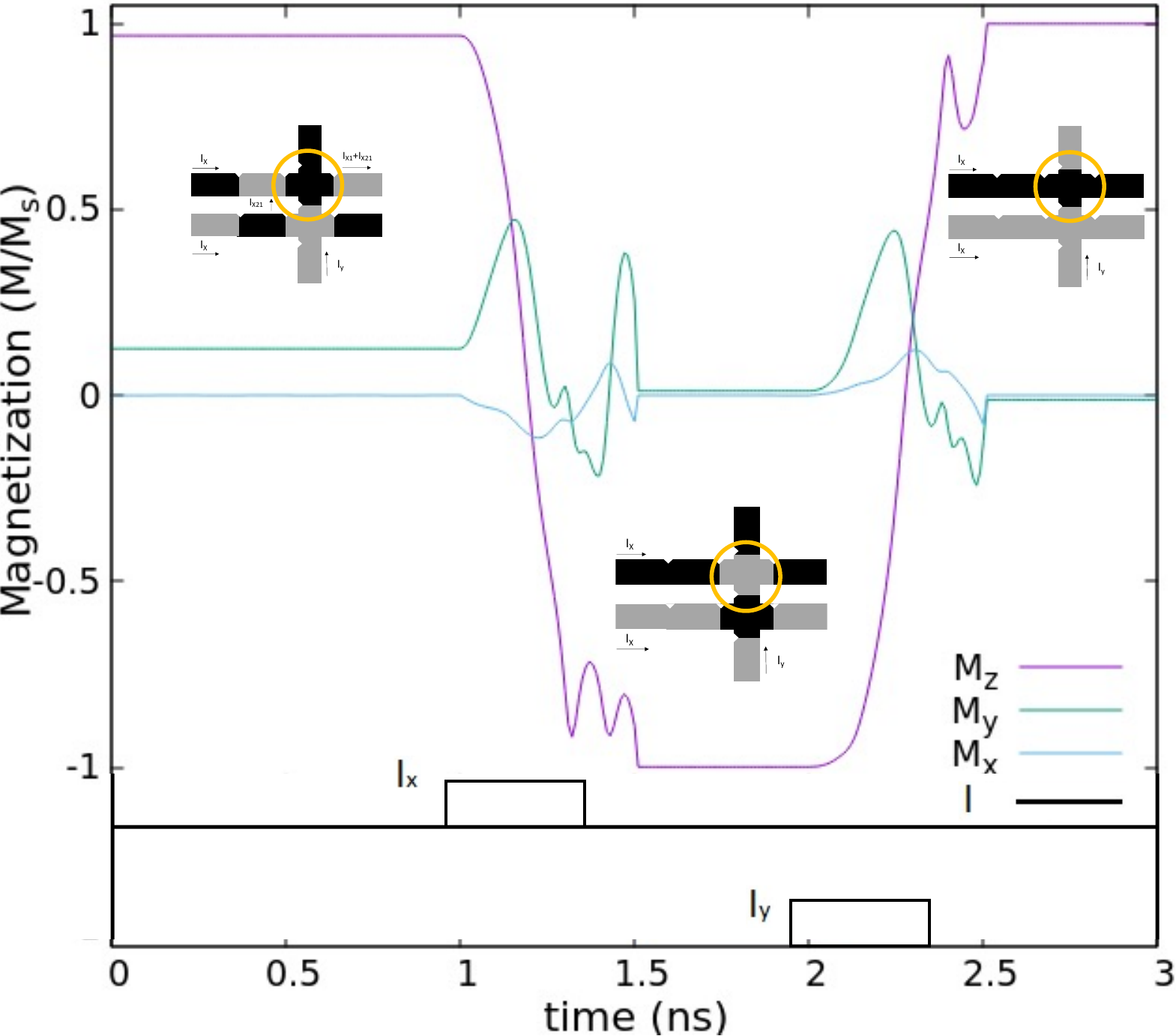}
\caption{Shift operation with Current}
\label{fig:X-cell-Mag}
\end{figure} 
Fig.~\ref{fig:X-cell-Mag} shows magnetization plots of the top X-cell (circled in gold) in a two X-NW bundle (each holding 4 domains) and a Y-NW  (two domains). Both Y-NW domains are cross-shaped and are at location three from the left extremities.  In each step, we show a representation of the magnetization of the bundle +Z in black and -Z in gray.  We performed a shift operation by applying the shift current $I_{shift}$ to the X-NWs: $I_{X1}=I_{X2}=I_{shift}$. 
The magnetization plot demonstrates the change in magnetization of the top X-cell, dropping from +Z to -Z due to the X-direction shift.  The state of the lower X-Cell changes from -Z to +Z as indicated in the bundle image.  Next, we inject $I_Y=I_{shift}$ to initiate a shift in the Y dimension as shown in Fig.~\ref{fig:X-cell-Mag}.  After the injection of Y shift current, the X-Cell switched back to +Z. This demonstrates that both X and Y shifting work correctly. Please also note the single-domain behavior of the X-Cell in the +/-Z direction with no non-transient X and Y component.

\begin{figure}[h]
\vspace{-0.1in}
\center     
\includegraphics[width=3in, height=1.0in]{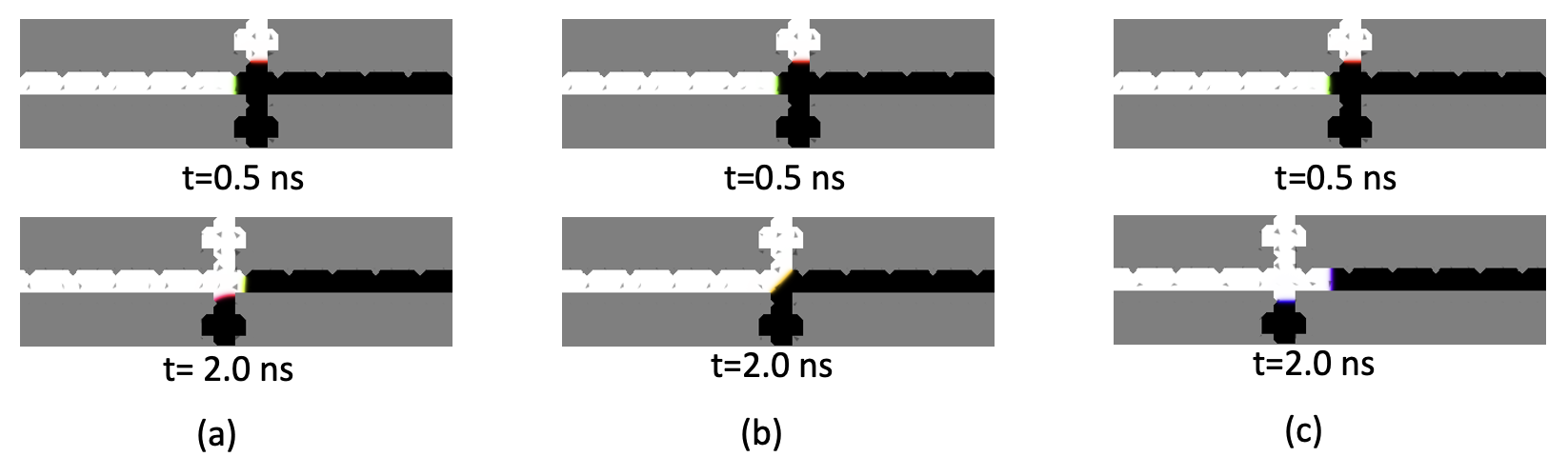}
\caption{Effect of different shift current.}
\label{fig:shift_current}
\vspace{-0.1in}
\end{figure}

{\bf Current-induced domain wall motion:} 
Here, we simulated a 2D nanowire with 8 domains in the X-NW and 3 domains in the Y-NW. The X-cell is placed in the middle. We have used spin transfer torque-induced DW motion. 
Fig.~\ref{fig:shift_current} shows DW motion in the X-NW with a thickness of 1nm and width of 40nm for different amounts of current density applied from the left side of the X-NW. 
During this shift operation, we made sure the current only flows through the X-NW by keeping the Y-NW in a high impedance path. 

Fig.~\ref{fig:shift_current} (a), (b), and (c) shows the effect of different shift  current density $J_{shift}$ on the DW motion. At t=0 ns, a DW was between the 4th and 5th domain of the X-NW and one DW located in the 1st and 2nd domain of the Y-NW. After t=0.5 ns a shift current is applied along the X-NW. For shift current density lower than $8\cdot 10^{11}$ A/m$^2$ the DW gets stuck between the corner of the cross Domain. And for shift current density over of $14\cdot10^{11}$ A/$m^{2}$ the DW gets over shifted. Thus, the upper and lower bound of the shift current density in the X-NW (Y-NW) was calculated to be $14\cdot 10^{11}$ ($12\cdot10^{11}$) and $8\cdot 10^{11}$ A/m$^2$ ($8\cdot 10^{11}$ A/m$^2$). Average X-NW shift current density $J_{avg}=(\lceil J_{shift} \rceil-\lfloor J_{shift} \rfloor)/2$ is 6.25\% higher than conventional X-NW without an X-cell.

\begin{figure}[h]
\begin{tabular}{c c}
\includegraphics[width=1.8in]{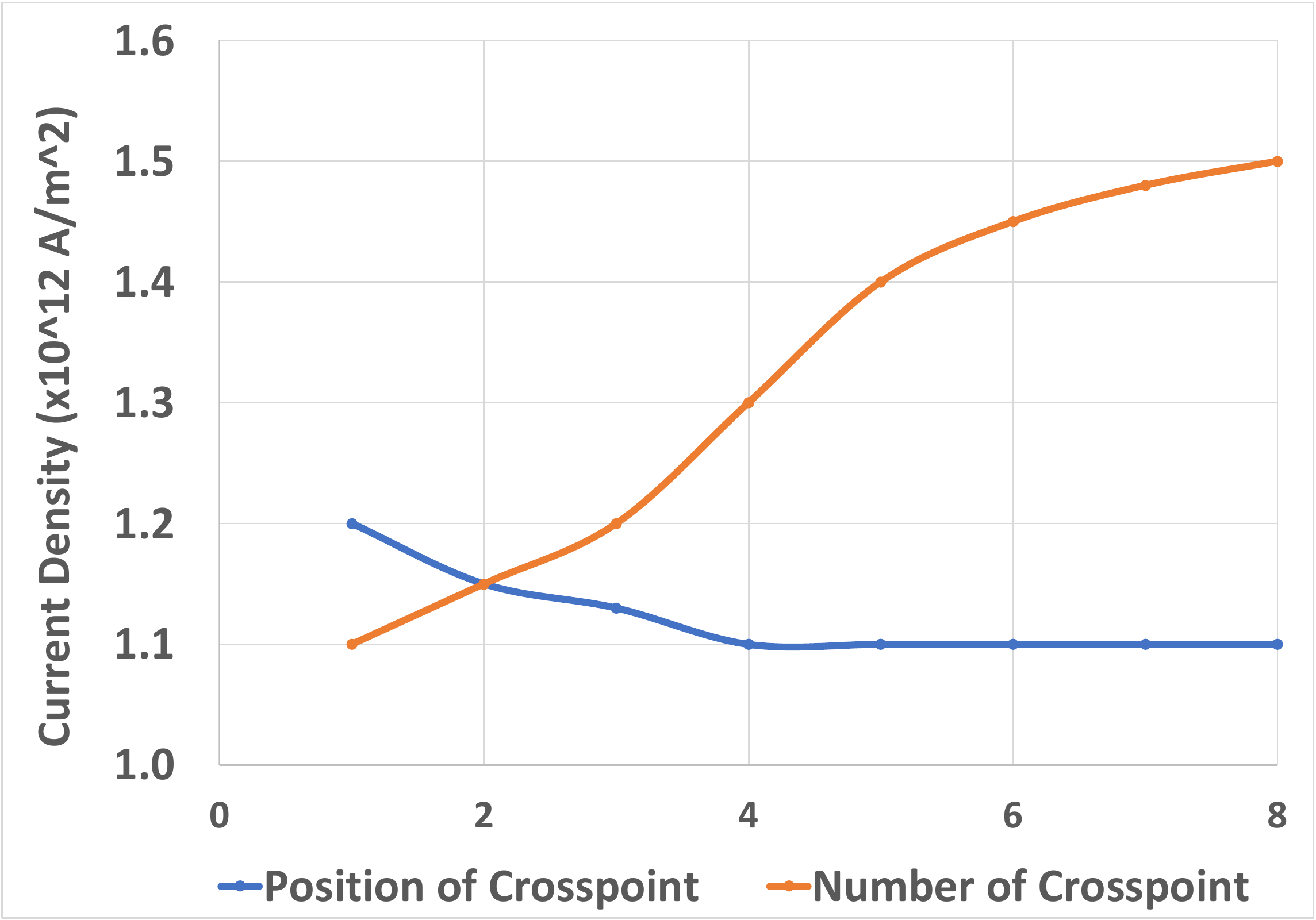}
&
\includegraphics[width=1.4in]{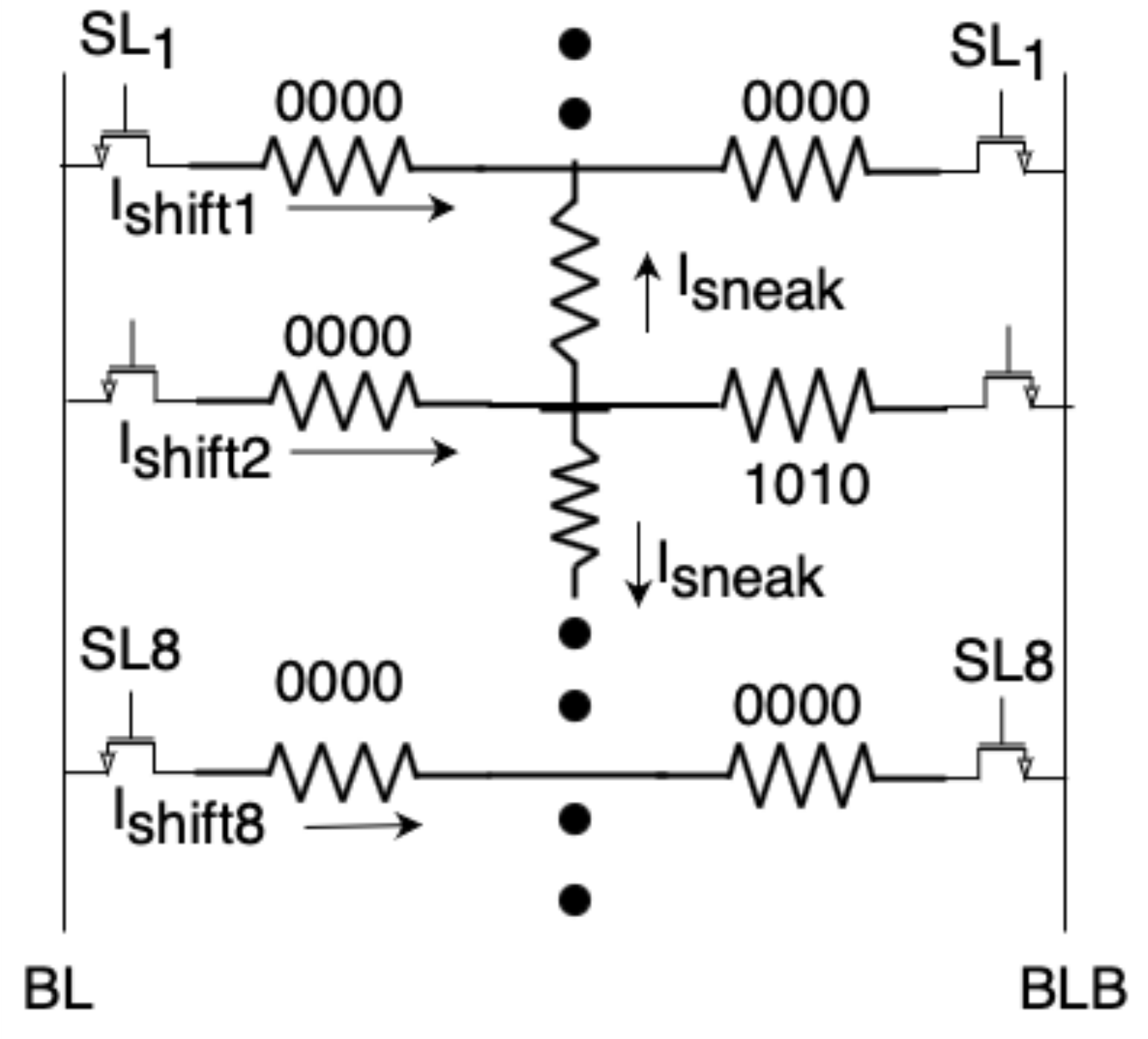}\\
(a) &(b)
\end{tabular}
\caption{(a) Plot of Shift current Density with Number of intersecting Y-NW and Placement of the X-Cell (only one intersecting Y-NW) in the X-NW. (b) Resistance Network analysis to analyze leakage current (sneak path) }
\label{fig:Performance}
\vspace{-0.1in}
\end{figure} 

{\bf Performance Analysis:} Fig.~\ref{fig:Performance} (a) (orange graph) shows that if the X-cell is near the shift current entry port the current requirement for shifting becomes higher. For other positions from middle to right port the required current is less and fairly stable.  Fig.~\ref{fig:Performance} (a) (blue graph) demonstrates that the shift current increases with the number of X-cell.

During shifting in the X-direction, the Y-NW is kept in a high impedance path by turning off the Y-NW access transistor. However, there will  still be some leakage through the Y-NW while we are sending current only through X-NWs because of the conductive cross-connection between them. Fig.~\ref{fig:Performance}b  shows the resistive network that models the leakage path that can exist between various X-NWs which are now connected via the Y-NW cells.   We have overlaid one 9-domain Y-NW in the middle of 8 bundled X-NWs. Since alternating `1's and `0's have the higher Anisotropic Magneto-Resistance (AMR) and homogeneous `0's or `1's have the lowest resistance, the worst-case leakage scenario will occur when only one path has the highest resistance while the others have the lowest resistance. In Fig.~\ref{fig:Performance}b, the second nanowire has the highest resistance path while all the other paths are low resistance.  All \texttt{SL} are turned ON and all \texttt{SLY} are turned OFF.  \texttt{BL} provides the shift current with $\overline{\text{\texttt{BL}}}$ is the sink to enable shifting. The maximum leakage current in this configuration is 1.08 uA (2.5\%) when only one nanowire is shifting in X-direction and  0.46uA (1.1\%) when all nanowires are shifting in X-direction. We scaled the study to 32 parallel X-NWs, which resulted in leakage currents of 2.3\% and 1.9\%, for shifting one and all X-NWs, respectively.  Further extending to overlaying 7 Y-NWs only nominally increases leakage current to 2.9\% and 3.1\%, respectively, for shifting one and all X-NWs.

According to~\cite{al2016geometrically} the speed of a DW in a rectangular NW can be expressed with the equation: $\mathbf{v}=\frac{\beta\gamma\hbar\mathbf{P}}{2\mathbf{e}\alpha\mathbf{M_s}}\mathbf{J_{shift}}$, where $\hbar$ is the reduced Planck constant. Our experiment shows the domain wall velocity in X- and Y- direction is $155$ m/s which aligns with the analytical calculation.Though the speed of a DW does not depend on the shape or NW geometry, the leakage current can affect the DW speed. Since the maximum leakage current is only 2.46\% in the worst-case scenario, the DW velocity is not impacted significantly.

In conclusion, we demonstrate the feasibility of cross-nanowires for 2D data shifting.  XDWM provides critical logical shifting that extends state-of-the-art processing in DWM.  In future work, we will explore XDWM feasibility and optimization for spin-orbit torque and other DWM enhancements.





\bibliographystyle{IEEEtran}
\bibliography{2D}
 \vspace*{-0.2in}
\end{document}